\documentstyle[twoside,fleqn,espcrc2,epsfig]{article}

\newcommand{\AmS}{{\protect\the\textfont2
  A\kern-.1667em\lower.5ex\hbox{M}\kern-.125emS}}

\let\q=\quad
\def\hnt{h_{\nu_\tau}}
\def\vs{{\it vs.}}
\def\etal{{\it et al.}}
\def\antibar#1{\overline{#1}}
\def\nubar{\antibar{\nu}}
\def\nutaubar{\antibar{\nu}_\tau}
\def\nutau{{\nu_\tau}}

\def\pizero{\pi^0}

\title{CLEO Results on Tau Michel Parameters
}

\author{Alan J.~Weinstein\address{California Institute of Technology,
        Pasadena, CA 91125}%
        }
       
\begin{document}

\begin{abstract}
We present measurements of the tau Michel Parameters
made by the CLEO experiment. Three different analyses are performed:
a spin-independent lepton spectrum analysis
and a second spin-dependent analysis 
using $\ell^\pm$ \vs\ $\pi^\mp\pi^0$ events,
and a third spin-dependent analysis using $\pi^+$ \vs\ $\pi^-$ events.
the results are used to derive limits on 
the general four-fermion couplings,
the mass of the charged Higgs in the MSSM, 
and a right-handed $W$ in left-right models.
Many of these measurements are more precise than the PDG
world averages.
\end{abstract}

\maketitle

\section{INTRODUCTION}

Heavy lepton ($\mu$, $\tau$) decays are 
mediated (in the Standard Model) by
the charged weak current, carried by $W^\pm$ force bosons
with a $V-A$ Lorentz structure.
As a consequence, such decays exhibit 
maximal parity violation, while conserving CP.
The couplings of all the charged leptons 
to the $W^\pm$ are consistent with being exactly equal
(universality).

Many extensions to the Standard Model predict 
small deviations from the $V-A$ law,
and/or fermion universality.
Such deviations can be observed 
in sensitive measurements of the Lorentz structure 
of the weak interaction.
In leptonic decays of the $\mu$ and $\tau$ leptons,
in which the outgoing neutrinos are unobserved,
the dynamics of the decay can be described
in terms of the Michel parameters \cite{ref:michel}.
The precise measurement of these parameters
can be used to search for evidence of, 
or provide limits on, processes which go beyond
the pure $V-A$ structure of the charged weak current.
In particular, they are sensitive to the presence
of small  {\it scalar} currents
(such as those mediated by the charged Higgs of the
Minimal Supersymmetric extensions \cite{ref:haber}
to the Standard Model, MSSM),
or small deviations from maximal parity violation 
(such as those mediated by the right-handed $W_R$ of 
left-right symmetric extensions \cite{ref:beg} 
to the Standard Model).

In the decays of the $\tau$ to $\ell\nu\overline{\nu}$, information on
the decay can be extracted from the shape of the momentum distribution
of the lepton $\ell$, and from its angular distribution relative to
the parent $\tau$ spin direction~\cite{ref:michel,ref:lspec}. After
integration over the unobserved neutrino momenta and the spin of
$\ell$, and neglecting radiative effects we can write the charged
lepton momentum spectrum as:  

\begin{eqnarray*}
\lefteqn{\frac{1}{\Gamma} \frac{d\Gamma}{dxd\cos\theta} 
       = \frac{x^2}{2} \times } \\
&&       \left[ \left( 12(1-x) + \frac{4{ \rho}}{3}(8x-6) 
           + 24{ \eta}{ \frac{m_\ell}{m_\tau}}
                \frac{(1-x)}{x}\right) \right. \\
     &   & \left. \pm { P_\tau }
               { \xi} { \cos\theta} \left( 4(1-x)+ 
        \frac{4}{3}{ \delta}(8x-6) \right)\right] \\
   && \propto  x^2\left[ I(x\vert {\rho , \eta} ) 
          \pm { P_\tau} A( x,{ \theta} \vert
            {\xi ,\delta}) \right]
\end{eqnarray*}

where $\rho$ and $\eta$ are the spectral shape Michel
parameters and $\xi$ and $\delta$ are the spin-dependent 
Michel parameters~\cite{ref:michel};
$x=E_{\ell}/E_{max}$ is the daughter charged lepton energy
scaled to the maximum energy $E_{max} = (m_{\tau}^2 +
m_{\ell}^2)/2m_{\tau}$ in the $\tau$ rest frame;
$\theta$ is the angle between the tau spin direction and the
daughter charged lepton momentum in the $\tau$ rest frame;
and $P_\tau$ is the polarization of the $\tau$.
In the Standard Model
(SM), the Michel Parameters have the values
$\rho=3/4$, $\eta=0$, $\xi = 1$ and $\delta = 3/4$.
Since $\tau^{+}\tau^{-}$ events are
produced with no net polarization at $e^{+}e^{-}$ center-of-mass
energies below the $Z^0$ mass, the lepton momentum spectrum alone
is not sensitive to the
spin-dependent parameters $\xi$ and $\delta$. 
These latter parameters are measurable by analyzing
the decays of both taus in an event.
In addition, semi-hadronic tau decays permit the measurement of 
the parameter $\hnt \equiv \xi_h = -g_v g_a /(g_v^2 + g_a^2)$,
sometimes referred to as the ``tau neutrino helicity'',
which takes the value $+1$ in the $V-A$ SM.

\section{CLEO DATA SAMPLE AND ANALYSIS}

Here we report on measurements of these Michel parameters
in three separate analyses performed on data from the CLEO II
experiment \cite{ref:CLEOII}
at Cornell's CESR collider, from the reaction
$e^+e^-\to\tau^+\tau^-$ at center of mass energy near 10.58 GeV.

In the first analysis, events of the type 
             { $\ell^-\nubar\nutau$} 
      \vs\ { $\pi^{+}\pizero\nutaubar$} are 
      selected\footnote{Throughout this paper,
                     charge conjugate processes are implied.}
      The { $\pi^\pm\pi^0$} system is used as a tag
      (to identify a tau pair event), and to estimate
      the tau direction in the lab.
      The lepton energy spectrum is measured and the 
      parameters { $\rho$} and { $\eta$}
      are extracted for $\tau\to e\nu\nu$ and $\tau\to\mu\nu\nu$.

In the second analysis, the same { $\ell^-\nubar\nutau$}
      \vs\ { $\pi^{+}\pizero\nutaubar$} sample used in the
      first analysis is again used.
      Here, the $\pi^\pm\pi^0$ system is used to analyze
      the spin of the decaying $\tau^\pm$.
      The spin correlations produce a correlation between the
       $\pi^\pm\pi^0$ system from one tau and the 
      momentum of the charged lepton from the other tau.
      A fitter which extracts all the available information
      from the full measured kinematics of these events
      is used to measure the parameters
      { $\rho$}, { $\xi$}, 
      { $\delta$}, and { $|\hnt|$}.
      The value of $\rho$ obtained in this analysis
      supersedes that of the first, and no attempt 
      is made to re-measure $\eta$.

In the third analysis, events of the type
      { $\pi^-\nutau$} \vs\ { $\pi^+\nutaubar$}
      are selected. The tau spin correlations
      produce correlations in the momenta of the two pions,
      which are used to extract $|\hnt|^2$.

The first and second analyses use $3.0\times 10^{6}$ produced
tau pairs, while the third uses only $1.5\times 10^{6}$.

For the first two analyses, we select
events where one $\tau$ decays leptonically ($e^\mp$ or $\mu^\mp$),
and the other decays to { $\pi^\pm\pizero\nutau$}.
The $\pi^\pm\pizero\nutau$ mode is used because
it has a large branching fraction,
negligible background from 
$e^+e^-\to e^+e^-$, $\mu^+\mu^-$, or $q\overline{q}$,
and a high, well-understood trigger efficiency.

\section{LEPTON SPECTRUM ANALYSIS}

In the first analysis~\cite{ref:mandeepa},
the $\rho^\mp \to\pi^\mp\pi^0$ decay is used to
estimate the $\tau^\mp$ flight direction.
The angle between the $\rho^\mp$ momentum
and the tau momentum in the lab can be calculated
(in the absence of radiative effects).
In approximately 60\%\ of the events,
this angle is less than 11$^\circ$
and is used to estimate the $\tau^\mp$ direction,
and thus the recoiling $\tau^\pm$ direction
in the lab. This permits us to boost the daughter
charged lepton $\ell^\pm$ momentum into the tau rest frame
so estimated (the ``pseudo-rest frame'' \cite{ref:Argus}).
For events in which the above angle is greater
than 11$^\circ$, the $\ell^\pm$ momentum is analyzed in the lab frame.

Muons with lab frame momentum $p_\mu < 1.5$ GeV/c
are not well identified in the CLEO muon detectors.
We can select decays with one charged track
which is inconsistent with being an electron,
and which has no extra photons from $\pi^0$ decays.
Such decays are most likely either $\tau\to \mu\nu\nu$
or $\tau\to \pi\nu$.
If the energy of the charged track 
in the pseudo-rest frame $E_\mu < 0.6 m_\tau$
then the event is kinematically inconsistent with being
a $\tau\to \pi\nu$ decay. 
We thus obtain a sample of $\tau\to \mu\nu\nu$ decays
with low momentum muons (particularly sensitive to the
$\eta$ parameter), with an estimated purity of 96\%.

The final $\tau\to e\nu\nu$ decay sample consists of $18587$ 
events analyzed in the pseudo-rest frame and 12981
analyzed in the lab frame.
The final $\tau\to \mu\nu\nu$ decay sample consists of 12580
events with muons identified in the CLEO muon detector
and 2931 events with muons identified kinematically
and analyzed in the pseudo-rest frame,
and 9186 events with muons identified in the CLEO muon detector
and analyzed in the lab frame.
Backgrounds from pions misidentified as hadrons
are measured from the data and subtracted.
All other backgrounds are estimated by the Monte Carlo
to be negligible.

\begin{figure}[hbt]
  \centering
  \psfig{figure=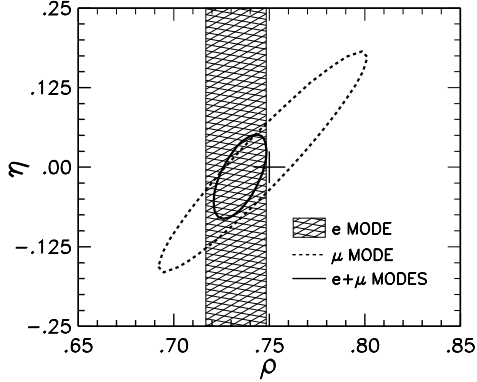,width=2.6in}
  \caption[]{Results from the lepton spectrum analysis.
            The shaded band denotes the electron 
             mode result, the large dotted contour indicates the muon mode 1$\sigma$ 
             error ellipse obtained in the $\eta-\rho$ plane,
             and the small dark ellipse indicates the 
             corresponding 1$\sigma$ result obtained from the
             simultaneous fit to both modes.
             The Standard Model expectation is indicated by the cross.}
  \label{fig:mand3}
\end{figure}

The resulting momentum spectra were fitted to Monte Carlo \cite{ref:koralb}
distributions, including all radiative effects, $\rho^\mp$ dynamics,
spin correlations, and detector efficiency and resolution.
The two electron spectra were fitted with one parameter, $\rho_e$,
with the result  $\rho_e =  0.732\pm0.015$.
The three muon spectra were fitted with two parameters, 
$\rho_\mu$ and $\eta_\mu$,
with the results
$\rho_\mu  =  0.747\pm0.055$ and 
$\eta_\mu  =  0.010\pm0.174$;
these values are strongly correlated (see Fig.~\ref{fig:mand3}).
All five spectra were fitted simultaneously, 
with the constraint that $\rho_e = \rho_\mu \equiv \rho_{e\mu}$,
with the two parameters $\rho_{e\mu}$ and $\eta_{e\mu}$.
Note, here, that $\eta_{e\mu}$ is the value of $\eta$ appropriate
for the muon decays, since the electron decays
are largely insensitive to the value of $\eta$.
The subscripted $\eta_{e\mu}$ is simply a reminder that
the result is obtained with the $\rho_e = \rho_\mu$ constraint.
The results are:
$\rho_{e\mu}=  0.735\pm0.014$ and $\eta_{e\mu} = -0.015\pm0.066$.

\begin{figure}[hbt]
  \centering
\psfig{figure=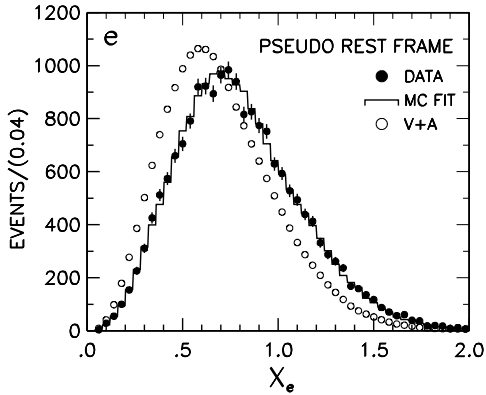,width=2.6in}
    \caption[]{The electron scaled pseudo rest frame energy spectra
               (solid points are data, histogram is the fit function).
             Open circles give the MC predicted spectrum for $V+A$.
             Events with $X>1$ result from the imperfect 
             reconstruction of the $\tau$ direction.}
  \label{fig:mand1}
\end{figure}

\begin{figure}[hbt]
  \centering
  \psfig{figure=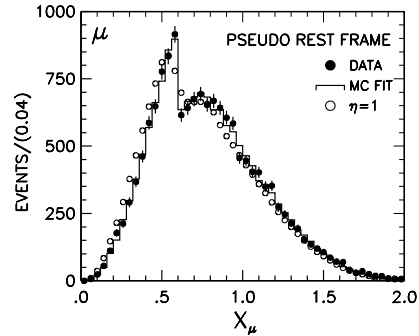,width=2.2in}
    \caption[]{The muon scaled pseudo rest frame energy spectra
               (solid points are data, histogram is the fit function).
             Open circles give the MC predicted spectrum for $\eta = 1$.
             The addition of the low momentum muons results in 
             the discontinuity observed at $X_{\mu}=0.6$.} 
  \label{fig:mand2}
\end{figure}

The spectra in the pseudo-rest frame,
and the fit results, are shown in Figs.~\ref{fig:mand1}
and \ref{fig:mand2}.
the results are summarized in Fig.~\ref{fig:mand3}.
Many systematic studies and 
fit variations were performed, 
all consistent with each other.
The results quoted above include systematic errors.
All these results are consistent with the Standard Model 
expectation that $\rho_e = \rho_\mu = 3/4$ and $\eta_\mu = 0$.

\section{SPIN-DEPENDENT ANALYSIS}

The second analysis~\cite{ref:moritz}
uses the same event sample as in the first,
{ $\ell^-\nubar\nutau$} \vs\ { $\pi^{+}\pizero\nutaubar$},
but the kinematically-identified low-momentum muons are not used.
We analyze the $\tau^+\to \pi^+\pi^0\bar{\nu}_\tau$ decay
to extract information on the $\tau^+$ spin orientation;
the QED-predicted spin correlations in $e^+e^-\to\tau^+\tau^-$
thus give information on the recoil $\tau^-$ spin direction.
We then analyze the momentum of the charged lepton in the
$\tau^-\to\ell^-\nubar_\ell\nu-\tau$ decay
in order to extract { $\xi$} and { $\delta$},
separately for electronic and muonic decays.
At the same time, we measure the magnitude of the 
tau neutrino helicity $\hnt$, and re-measure 
the $\rho$ parameter.
We assume the value of $\eta$ as measured from the 
first analysis.
We also select 11177 events of the type
{ $\pi^{-}\pizero\nutau$} \vs\ { $\pi^{+}\pizero\nutaubar$},
to extract $|\hnt|$ with greater precision.

The $\rho^+\to\pi^{+}\pizero$ decay is used to analyze
the parent tau spin orientation. The energy of the 
$\pi^{+}\pizero$ system, and the angle of the pion momentum
in the $\pi^{+}\pizero$ rest frame relative to the
$\pi^{+}\pizero$ momentum in the lab frame can be combined
to give a ``polarimeter'' variable~\cite{ref:davier}
$\omega = \left< \vec{p}_\tau \cdot \vec{H_\tau} \right>_{\phi_\nutau}
   = \omega(E_\rho, \cos\theta(\rho\to\pi\pi))$.
Clear correlations between the value of $\omega$ from
the $\tau^+$ decay and the charged lepton momentum from 
the $\tau^-$ decay are observed, consistent with Standard Model
expectations.

To make full use of all the available kinematical information, 
we do a multi-dimensional likelihood fit to all of the 
observed three-momenta in the lab,
using the full production and decay matrix element,
including the effects of radiation (ISR/FSR, decay, 
and external bremsstrahlung).
Schematically, for the $\mu-\vs\-\rho$ sample,
\begin{eqnarray*}
{\vert {\cal M} \vert}^2 
  & = & H_1 \times { P}  \times [L_1 + {\rho} L_2 + {\eta}   L_3  ] \\
  &   & + { h_{\nu_\tau}} H^\prime_{1\alpha} 
       \times  { C^{\alpha\beta}} \times [ {\xi} L^\prime_{1\beta} +
  {\xi\delta} L^\prime_{2\beta} ] .
\end{eqnarray*}
Here, $H_1$ describes the spin-independent part of the 
$\tau\to\pi\pi^0\nu$ decay rate,
$L_1$ through $L_3$ describe the spin-independent part of the 
$\tau\to\ell\nu\nu$ decay rate (with explicit Michel parameter dependence),
and $P$ describes the spin-independent part of the
$\tau^+\tau^-$ production.
$H^\prime_{1\alpha}$, $L^\prime_{1\beta}$ and $L^\prime_{2\beta}$,
and $C^{\alpha\beta}$ describe the analogous 
spin-dependent quantities. All are functions of the momenta
of all the observed particles.

Evaluation of the likelihood per event requires 
integration over the unobserved neutrinos and radiated photons:
a 22 dimensional integral. This is performed 
using a unique ``reverse Monte Carlo'' technique.
All major ``backgrounds''
($\pi, K\pi, \pi2\pi^0 \to \pi\pi^0$, $\pi\to \mu$) 
are modeled, with full Michel parameter dependence.
The detector efficiency and resolution is fully modeled.
The fit procedure was tested with many ensembles of Monte Carlo events
with all three topologies ($e-\vs\-\rho$, $\mu-\vs\-\rho$, 
$\rho-\vs-\rho$), including backgrounds;
no bias was observed.

A measurement of the spin-dependent Michel parameters
allows one to distinguish the Standard Model $V-A$ interaction
(left-handed $\nu_\tau$) from $V+A$ (right-handed $\nu_\tau$).
The probability that a right-handed (massless) tau neutrino
participates in the decay can be expressed as
$$P^\tau_R = 1/2 \left[1+ 1/9 \left(3 {\xi}
              -16{\xi\delta} \right)\right], $$
and $P^\tau_R = 0$ for the SM $V-A$ interaction.

\begin{figure}[th]
 \psfig{figure=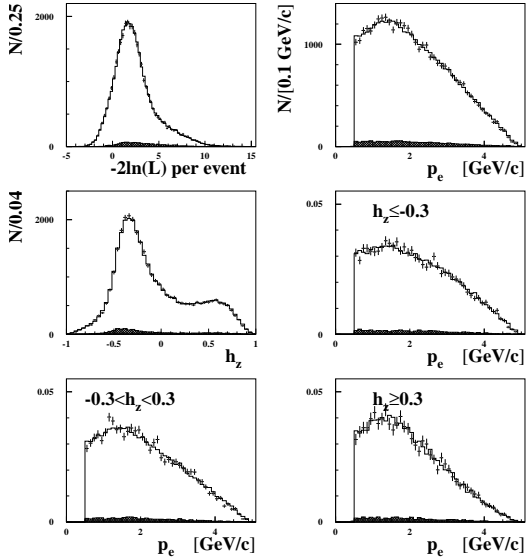,width=3.0in}
  \caption[]{Comparison of the $e-\vs\-\rho$ data (points)
    with the fit results (histograms) for the following quantities:
    the $-2\ln{\cal L}$ (log-likelihood) per event;
    the electron momentum in the lab;
    the polarimeter variable $\omega\equiv h_Z$ for the $\rho$ decay,
    and  the electron momentum for three 
    ranges of polarimeter variable of the recoiling $\rho$ decay:
    $\omega \le -0.3$, $-0.3 < \omega < 0.3$, and $\omega \ge 0.3$.
}
  \label{fig:morkin}
\end{figure}

The results of the fits to the $e-\vs\-\rho$ sample are: 
\begin{eqnarray*}
\rho_e &=& 0.747\pm 0.012\pm 0.004 \q\q (SM = 3/4) \\
\xi_e &=& 0.979 \pm 0.048 \pm 0.016 \q\q (SM = 1) \\
\xi_e \delta_e &=& 0.720 \pm 0.032 \pm 0.010 \q\q (SM = 3/4). 
\end{eqnarray*}
All measures of goodness-of-fit yield satisfactory results.
From the distribution of the likelihoods for each event
as compared to Monte Carlo expectations,
the confidence level (CL) for the fit is 73\%.
Comparisons of various kinematical quantities
between the data and the results of the fit
are shown in Fig.~\ref{fig:morkin}.
We extract the limit $P_\tau^R < 0.065$ at $90\%$ CL.
The last three plots in this figure
clearly demonstrate the spin correlations
and the effectiveness of the fit in modeling them.

For the $\mu-\vs\-\rho$ sample, the results (using 
$\eta_\mu = 0.010$ from the first analysis) are:
\begin{eqnarray*}
\rho_\mu &=& 0.750\pm 0.017 \pm 0.045 \\
\xi_\mu &=& 1.054 \pm 0.069 \pm 0.047  \\
\xi_\mu \delta_\mu &=& 0.786 \pm 0.041 \pm 0.032 ,
\end{eqnarray*}
with a confidence level for the fit of 21\%.
We extract the limit $P_\tau^R < 0.067$ at $90\%$ CL.

For the $\rho-\vs\-\rho$ sample, the results are: 
\begin{eqnarray*}
h_{\nu_\tau}^2  &=& 0.995 \pm 0.010 \pm 0.003 \q\q (SM = 1)
\end{eqnarray*}
with a confidence level for the fit of 9\%.

The following sources of systematic errors were considered:
Monte Carlo statistics;
lepton identification efficiency;
spin analyzer efficiency;
modeled backgrounds;
remaining unmodeled backgrounds;
contribution of  $\rho^\prime$ to the $\pi\pi^0$ dynamics;
the value of the Michel parameter $\eta$;
detector energy, momentum, and angular resolution;
and modeled and unmodeled radiation effects.
There is no sign of significant systematic bias,
and Monte Carlo statistics remains the largest source
of systematic error in most cases.

Under the assumption of lepton universality,
we can constrain $\rho_e = \rho_\mu \equiv \rho_{e\mu}$,
and similarly for $\xi$ and $\xi\delta$. We then obtain: 
\begin{eqnarray*}
\rho_{e\mu} &=& 0.747\pm 0.010 \pm 0.006 \\
\xi_{e\mu} &=& 1.007 \pm 0.040 \pm 0.015  \\
(\xi\delta)_{e\mu} &=& 0.745 \pm 0.026 \pm 0.009 . 
\end{eqnarray*}
We extract the limit $P_\tau^R < 0.044$ at $90\%$ CL.

These results are consistent with the Standard Model Prediction,
and are, in most cases, more precise than the world average
values ~\cite{ref:pdg96}.
These results supersede those on $\rho_e$, $\rho_\mu$, $\rho_{e\mu}$ 
from the first analysis.

\section{SPIN CORRELATIONS IN $\pi^+$ \vs\ $\pi^-$}

In the third analysis \cite{ref:masui},
we again exploit the fact that
in $e^+e^-\to\gamma^*\to\tau^+\tau^-$ at 10 GeV,
the $\tau$ spins are 95\% anti-correlated.
If both taus decay semi-hadronically, then
in the lab frame, the differential distribution of the
energies of the two hadronic systems
scaled to the beam energy, $x_\pm \equiv E_\pm/E_b$
is given by:
$$ \frac{d^2\sigma}{d{ x_+} d{ x_-}} = 
   F_1({ x_+},{ x_-}) +
   { |\hnt|^2}  F_2({ x_+},{ x_-})
$$
where the functions $F_1$ and $F_2$ are well known and are
particularly simple for $\pi^+$ \vs\ $\pi^-$ events.

\begin{figure}[bht]
  \centering
  \psfig{figure=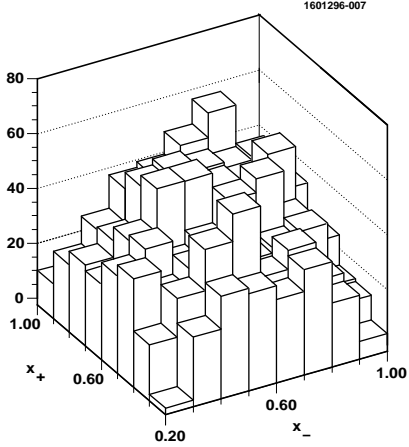,width=2.2in}
\caption[]{
The raw correlation between the scaled energies $x_\pm \equiv E_\pm/E_b$
of the two pions in 
($\tau^-\to\pi^-\nutau$) \vs\ ($\tau^+\to\pi^+\nutaubar$) events.}
\label{fig:masui}
\end{figure}

We select events of the type
($\tau^-\to\pi^-\nutau$) \vs\ ($\tau^+\to\pi^+\nutaubar$),
with no extra photons.
In order to suppress $\mu \to \pi$ backgrounds,
we require the pions to shower in the calorimeter.
We use our limited dE/dx separation to suppress
$\tau\to K^\pm\nu$.
Kinematical cuts suppress two-photon events
and other backgrounds.
From $1.5\times 10^{6}$ produced tau pairs,
we obtain 2041 $\pi^+$ \vs\ $\pi^-$ candidates.
The distribution of scaled energies
$x_+$ \vs\ $x_-$ is given in Fig.~\ref{fig:masui}.

From a fit to this distribution (and related ones),
we obtain $|\hnt| = 1.03\pm0.06\pm0.04$,
consistent with the SM expectation of 1.

\section{DISCUSSION OF RESULTS}

\begin{figure}[bht]
  \centering
   \psfig{figure=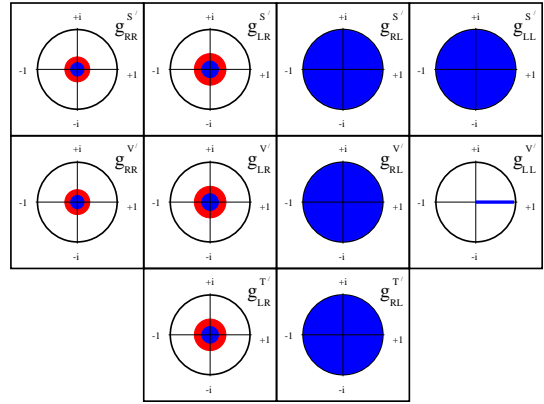,width=2.8in}
    \caption[]{
$90\%$ confidence limits on the reduced coupling constants 
$g^{\gamma^\prime}_{\epsilon\mu}= 
g^\gamma_{\epsilon\mu}/max(g^\gamma_{\epsilon\mu})$,
in the complex plane.
The circles with unit radii indicate the maximum possible values, the
other solid circles indicate the constraints obtained using present
world average results, and the hatched regions indicate the new 
constraints obtained after including the results presented here.}
\label{fig:couple}
\end{figure}

From these results, limits can be obtained on the
complex couplings in the 
generalized four-fermion interaction model \cite{ref:michel,ref:lspec};
these are shown in Fig.~\ref{fig:couple}.
the couplings are shown normalized to their maximum value
(hence the prime$^{\prime}$),
and $g_{LL}^{V\prime}$ is fixed to be real.
The new results improve the limits on couplings
to right-handed tau neutrinos, but are unable to 
distinguish amongst the couplings to left-handed neutrinos
without independent information such as 
a measurement of the cross-section $\sigma(\nu_\tau e \rightarrow \tau\nu_e)$.
In the $V-A$ Standard Model, all of the couplings are zero
except $g_{LL}^{V\prime} = 1$.
The results are clearly consistent with the Standard Model,
but are still at the level of precision
achieved for muon decays.

\subsection{Limits on Charged Higgs}

In the minimal supersymmetric extension to the 
Standard Model (MSSM), a charged Higgs boson
will contribute to the decay of the tau,
interfering with the left-handed $W^-$ diagram,
and producing a non-zero value for $\eta$.
For $\tau\to\mu\nu\nu$, 
$$\eta = - \;\;\frac{m_{\tau} m_{\mu} tan^{2}\beta}{2 m^{2}_{H}},$$
and similar formulas exist for ${ \xi}$ and ${ \xi\delta}$.
With the measurements reported here 
of $\eta$,  $\xi$, and $\xi\delta$ combined,
we can extract the limit
$m_{H^\pm} > 1.02 \tan\beta\mbox{ GeV/c}^2$ at  90\% CL.
This limit is significant in comparison with 
\def\rapprox{\sim\kern-1em\raise 0.6ex\hbox{$>$}}
direct search limits, in the region $\tan\beta \rapprox 200$
\cite{ref:stahl}.

\subsection{$W_R$ in Left-right symmetric models}

In left-right symmetric models \cite{ref:beg},
there are two sets of weak charged bosons $W^\pm_1$ and $W^\pm_2$,
which mix to form the observed ``light''
left-handed $W^\pm_L$ and a heavier (so far, unobserved)
right-handed $W^\pm_R$.
The parameters in these models are
$\alpha = M(W_1)/M(W_2)$ (= 0 in the SM), and
$\zeta = $ mixing angle, $= 0$ in the SM.
The heavy right-handed $W^\pm_R$
will contribute to the decay of the tau,
interfering with the left-handed $W^-$ diagram,
and producing deviations from the Standard Model values
for the Michel parameters $\rho$ and $\xi$.

\begin{figure}[bht]
  \centering
    \psfig{figure=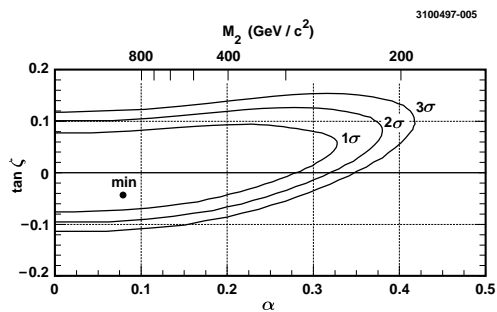,width=2.6in}
    \caption[]{Limits on the mass ratio $\alpha$ and the mixing angle
      $\zeta$ of a left-right symmetric model.}
  \label{fig:WLR}
\end{figure}

By using the results of these analyses assuming lepton universality,
especially the limit on $P^\tau_R$ 
(the probability that a right-handed tau neutrino
participates in the decay), we can obtain limits on
$\alpha$ and $\beta$ in these models,
shown in Fig.~\ref{fig:WLR}.
For mixing angle $\zeta = 0$, we obtain
$m_R > 304\mbox{ GeV/c}^2$ at 90 \% CL, and 
for free mixing angle $\zeta$, we obtain
$m_2 > 260\mbox{ GeV/c}^2$ at 90 \% CL.

\section{SUMMARY}

Using data from the reaction
$e^+e^-\to\tau^+\tau^-$ at center of mass energy near 10.58 GeV
collected with the CLEO II detector,
we have studied the Lorentz structure of the $\tau - W - \nu_\tau$ couplings.
we use three different analyses to extract measurements of 
the Michel Parameters $\rho$, $\eta$, $\xi$, $\delta$,
and the tau neutrino helicity $\hnt$.
All of the measurements 
are consistent with the predictions from the pure $V-A$ Standard Model,
at the $\sim$ few \%\ level.
They are, in most cases, more precise than the previous world averages.

From these measurements, we extract limits 
on the non-Standard Model complex couplings in the
generalized four-fermion interaction model.
We set a limit  on the mass of the charged Higgs in MSSM
models, significant for large $\tan\beta$.
We set limits on the mass and mixing angle
of a heavy right-handed $W$ predicted
in left-right symmetric models.

In addition, we also present at this conference
a precision measurement of the parity violating {\it sign} of $\hnt$
in $\tau\to 3\pi\nutau$ decay \cite{ref:schmidtler}.

Our results are still statistics limited,
and CLEO II already has 3 times more data collected
and now available for analysis.
It is worth our while to continue improving the precision of
these measurements, to search for small effects
due to high energy scale physics.

\section{ACKNOWLEDGEMENTS}

We gratefully acknowledge the effort of the CESR staff in providing us with 
excellent luminosity and running conditions. 
This work was supported by the U.S.~NSF and DoE.
I thank the conference organizers for a thoroughly stimulating 
and pleasant conference.

\end{document}